\documentclass[showpacs,preprintnumbers,amssymb,twocolumn]{revtex4-2}

\expandafter\let\csname equation*\endcsname\relax

\expandafter\let\csname endequation*\endcsname\relax
\usepackage{graphicx}
\usepackage{dcolumn}
\usepackage{color}
\usepackage{bm}
\usepackage{amsmath}
\usepackage{amssymb}
\usepackage{epsfig}
\usepackage{amsfonts}
\usepackage{lineno,hyperref}
\usepackage{array}
\usepackage{float}
\usepackage{microtype}
\usepackage{multirow}
\usepackage{adjustbox}
\usepackage[english]{babel}
\usepackage{epstopdf}
\usepackage{parskip}
\usepackage{subcaption}

\def \a{\alpha}
\def \b{\beta}
\def \l{\lambda}

\def \g{\gamma}

\def \k{\kappa}
\def \o{\omega}

\def \t{\theta}

\def \be{\begin{equation}}
\def \ee{\end{equation}}
\def \ben{\begin{eqnarray}}
\def \een{\end{eqnarray}}

\def \G{\bar{G}}
\def \R{\bar{R}}

\def \T{\bar{T}}

\def \La{\mathcal{L}}
\def \nb{\nabla}
\begin{document}
\title{Form Invariance of Raychaudhuri equation in the presence of Inflaton-type fields}

\author{Arijit Panda}
\email{arijitpanda260195@gmail.com}
\affiliation{Department of Physics, Raiganj University, Raiganj, Uttar Dinajpur, West Bengal 733 134, India.  $\&$\\
Department of Physics, Prabhat Kumar College, Contai, Purba Medinipur 721 404, India.}

\author{Debashis Gangopadhyay}
\email{debashis.g@snuniv.ac.in}
\affiliation{Department of Physics, School of Natural Sciences, Sister Nivedita University, DG 1/2, Action Area 1 Newtown, Kolkata 700156, India. }

\author{Goutam Manna$^a$}
\email{goutammanna.pkc@gmail.com\\$^a$Corresponding author}
\affiliation{Department of Physics, Prabhat Kumar College, Contai, Purba Medinipur 721404, India $\&$\\ Institute of Astronomy Space and Earth Science, Kolkata 700054, India}

\date{\today}
  
\begin{abstract}
We show that the Raychaudhuri equation remains form invariant for certain solutions of scalar fields $\phi$ whose Lagrangian is non-canonical and of the form $\mathcal{L}(X,\phi)=-V(\phi)F(X)$, with $X=\frac{1}{2} g_{\mu\nu} \nabla^{\mu}\phi \nabla^{\nu} \phi$ and $V(\phi)$ the potential. Solutions exist for both homogeneous and inhomogeneous fields that are like inflatons. Certain recent observations indicate that the cosmos is inhomogeneous and thus  our results are in sync with latest observations. So the Raychaudhuri equation can accommodate primordial inhomogeneities as well as cosmologically relevant scenarios.

\end{abstract}



\maketitle

\section{Introduction:} 
This article aims to demonstrate the form invariance of the Raychaudhuri equation (RE) ~\cite{Ray52,Ray53,Ray55,Ray57,Ray,Kar} in the presence of both homogeneous and inhomogeneous scalar fields whose Lagrangian is non-canonical and associated with K-essence geometry. The aforementioned fields exhibit conditional correspondences with inflaton fields, which persist over the whole of the universe's early to late phases. So the Raychaudhuri equation may also account for primordial inhomogeneities.  We focus on the Raychaudhuri equation here because this is a geometric equation that has fundamental significance without relying on any theory of gravity. Cosmological consequences have been discussed elsewhere for homogeneous fields~\cite{Das}. 

The Raychaudhuri equation describes the time-dependent characteristics of the congruence of geodesics as witnessed by another nearby congruence of geodesics throughout its development. The evolution primarily describes the progression of congruence inside a specific spacetime framework and this progression is produced by a curve that follows the tangent vectors (which can be either timelike or null). The equation describes the rate at which the expansion of the congruence (a measurement not dependent on coordinate choices) is related to the gravitational influence.
It also illustrates the attractive nature of gravity.  Subsequently it emerged as a crucial lemma for investigating the renowned Hawking-Penrose singularity theorem\cite{Penrose65,Hawking65,Hawking66,Senovilla}. At a more fundamental level, the Raychaudhuri equation takes into account the anisotropy (regarding rotation) and inhomogeneity (regarding shear) of the cosmos \cite{Kar}. The equation gives the expansion rates of two null or timelike geodesics, which can converge or diverge depending on the energy conditions provided by the Ricci tensor. Converging geodesics can be used to explain the focusing theorem \cite{Poisson,Das}, while diverging geodesics can help to understand the non-focusing theorem due to the expansion of the universe \cite{Das,Choudhury}. Burger et al. \cite{Burger} have discussed the relevance and several uses of the Raychaudhuri equation beyond general relativity.

The plan of this paper is as follows: in Section 2 we briefly review the theory of K-essence which involves the non-canonical Lagrangian, Section 3 discusses the Raychaudhuri equation in the presence of K-essence, section 4 deals with the homogeneous and inhomogeneous solutions of the K-essence fields that leave the Raychaudhuri equation invariant, Section 5 relates the solutions to inflaton fields while Section 6 is the conclusion.

\section{K-essence}
An example of a non-canonical theory that investigates several avenues in cosmological research is the K-essence theory \cite{Das,Vikman,Picon1,Picon2,dg1,dg2,dg3,Chimento,Scherrer,Dutta,Santiago,Mukohyama,Babichev1,Visser}. There are different ways to write the Lagrangian of the K-essence field. One way is $\La(X,\phi)=-V(\phi)F(X)$ \cite{Vikman,Picon1,Picon2,dg1,dg2,dg3}, another is $\La (X,\phi)=F(X)-V(\phi)$ \cite{Dutta,Santiago}, and a third is $\La (X,\phi)\equiv \La (X)= F(X)$ \cite{Scherrer,Mukohyama}. Here, $F(X)\equiv \La(X) (\neq X)$ is the non-canonical kinetic part with $X=\frac{1}{2} g_{\mu\nu} \nabla^{\mu}\phi \nabla^{\nu} \phi$, $V(\phi)$ is the canonical potential part.  This theory emphasizes that the kinetic element dominates the potential part and shows a dynamic behavior that assures late-time acceleration without fine-tuning. Another noteworthy part of the K-essence idea is its capacity to create dark energy with a sound speed ($c_s$) continuously lower than light. This property may lessen cosmic microwave background disruptions at large angular scales. In addition, according to Scherrer \cite{Scherrer}, this theory may be applied to both dark energy and dark matter to form a unified theory. The Dirac-Born-Infeld (DBI) type non-canonical action \cite{Born,Dirac} was utilized by Manna et al. \cite{gm1,gm2,gm3,gm4,gm5,gm6,gm7,gm8}, and they have shown that the K-essence theory can be applied to the models of dark energy \cite{gm1,gm2,gm6} as well as in gravitational perspectives \cite{gm3,gm4,gm7,gm8}, without considering the dark sectors of the universe. The K-essence theory has also been utilized in describing the inflationary behavior of the early universe \cite{Picon1,Mukhanov,Ferreira}.  The results from the Planck collaborations \cite{Planck1,Planck2,Planck3} have investigated the empirical evidence that supports the idea of K-essence with a DBI-type non-canonical Lagrangian, among other things. The dynamical solutions of the K-essence equation of motion distinguish K-essence theories with non-canonical kinetic terms from relativistic field theories with canonical terms. These demonstrate spontaneous Lorentz invariance breakdown and metric changes under perturbations. The K-essence field Lagrangian's theoretical formulation has non-canonical features that the metric demands. These features distribute perturbations across the evolving analogue spacetime, or curved spacetime. This can imitate the background metric creating late-time acceleration at the right time frame in the universe's evolution and so is relevant in  the 'cosmic coincidence' problem.

\section{ Modified Raychaudhuri equation in the context of K-essence:} The action of the K-essence geometry can be written as \cite{Picon1,Picon2,Babichev1,Vikman,Scherrer}
\ben
S_{k}[\phi,g_{\mu\nu}]= \int d^{4}x {\sqrt -g} \La(X,\phi),
\label{1}
\een
where $\La(X,\phi)$ represents the non-canonical Lagrangian with the canonical kinetic term $X=\frac{1}{2}g^{\mu\nu}\nabla_{\mu}\phi\nabla_{\nu}\phi$. Here, the standard gravitational metric $g_{\mu\nu}$ has minimal coupling with the K-essence scalar field $\phi$. 

The energy-momentum tensor is  \cite{Babichev1,Vikman}:
\ben
T_{\mu\nu}\equiv \frac{-2}{\sqrt {-g}}\frac{\delta S_{k}}{\delta g^{\mu\nu}}&=-2\frac{\partial \La}{\partial g^{\mu\nu}}+g_{\mu\nu}\La\nonumber\\
&=\La_{X}\nabla_{\mu}\phi\nabla_{\nu}\phi
+g_{\mu\nu}\La,
\label{2}
\een
where $\La_{\mathrm X}= \frac{d\La}{dX}\neq 0,~\La_{\mathrm XX}= \frac{d^{2}\La}{dX^{2}},
~\La_{\mathrm\phi}=\frac{d\La}{d\phi}$ and $\nabla_{\mu}$ is the covariant derivative defined with respect to the gravitational metric $g_{\mu\nu}$. 

A scalar field equation of motion (EOM) is \cite{Babichev1, Vikman, gm1, gm2}
\ben
-\frac{1}{\sqrt {-g}}\frac{\delta S_{k}}{\delta \phi}= \tilde{G}^{\mu\nu}\nabla_{\mu}\nabla_{\nu}\phi +2X\La_{X\phi}-\La_{\phi}=0,
\label{3}
\een
where the effective metric is \\
$\tilde{G}^{\mu\nu}\equiv \frac{c_{s}}{\La_{X}^{2}}[\La_{X} g^{\mu\nu} + \La_{XX} \nabla ^{\mu}\phi\nabla^{\nu}\phi]$
with $(1+ \frac{2X  \La_{XX}}{L_{X}}) > 0$ and $c_s^{2}(X,\phi)\equiv{(1+2X\frac{\La_{XX}} {\La_{X}})^{-1}}$.

Using a conformal transformation \cite{gm1,gm2,gm3,gm4} $\bar G_{\mu\nu}\equiv \frac{c_{s}}{\La_{X}}G_{\mu\nu}$, the inverse metric of $\tilde{G}^{\mu\nu}$ is
\ben
\bar{G}_{\mu\nu}=g_{\mu\nu}-\frac{\La_{XX}}{\La_{X}+2X\La_{XX}}\nabla_{\mu}\phi\nabla_{\nu}\phi.
\label{5}
\een
The corresponding geodesic equation is \cite{gm1,gm2,gm3,gm4}
\ben
\frac {d^{2}x^{\alpha}}{d\l^{2}} +  \bar\Gamma ^{\alpha}_{\mu\nu}\frac {dx^{\mu}}{d\l}\frac {dx^{\nu}}{d\l}=0, \label{6}
\een
where $\l$ is an affine parameter and the new connection coefficients ($\bar{\Gamma}^{\a}_{\mu\nu}$) is
\ben
\bar\Gamma ^{\alpha}_{\mu\nu}=\Gamma ^{\alpha}_{\mu\nu} -\frac {1}{2(1-2X)}\Big[\delta^{\alpha}_{\mu}\partial_{\nu}
+ \delta^{\alpha}_{\nu}\partial_{\mu}\Big]X.
\label{7}
\een
The covariant derivative associated with the K-essence geometry ($\bar{G}_{\mu\nu}$) is
\ben
D_{\mu}A^{\nu}=\partial_{\mu} A^{\nu}+\bar \Gamma^{\nu}_{\mu\l}A^{\l}
\label{8}
\een
The corresponding Einstein equation in this geometry is \cite{Das, Vikman, Panda}:
$ \bar{\mathcal{G}}_{\mu\nu}=\R_{\mu\nu}-\frac{1}{2}\bar{G}_{\mu\nu}\R=\k \T_{\mu\nu}$, 
where $\k=8\pi G$ is constant and $\R_{\mu\nu}$ is the Ricci tensor, $\R~ (=\R_{\mu\nu}\bar{G}^{\mu\nu})$ and $\T_{\mu\nu}$ is the energy-momentum tensor of this geometry. 

In K-essence geometry, the Raychaudhuri equation for a time-like vector field $v^{\b}$ is \cite{Das}
\ben
\frac{d\tilde{\Theta}}{d\bar{s}}+(D_\a v^\b)(D_\b v^\a)=-\bar{R}_{\g\b}v^\g v^\b
\label{9}
\een
with $\bar{s}$ is an affine parameter, $\tilde{\Theta}\equiv D_\a v^{\b}$ the scalar expansion, and $\bar{R}_{\g\b}$ the Ricci tensor of this geometry.
The new geometry is torsion-free and hence equation (\ref{9}) is generated from the commutation relationship of the covariant derivatives (\ref{8}) as
\ben
[D_\a\, ,D_\b]v^\mu=\bar{R}^\mu_{\g\a\b}v^\g \label{9.1}
\een
where $\bar{R}^\mu_{\g\a\b}$ is the Riemann tensor. Following \cite{Das}, we rewrite the modified Raychaudhuri equation as
\ben
&&\frac{d\tilde{\Theta}}{d\bar{s}}+\Big(2{\sigma}^2-2{\omega}^2+\frac{1}{3}\t^2\Big)-\frac{\theta}{(1-2X)}v^\mu\partial_\mu X\nonumber\\
&&+\frac{7(v^\mu\partial_\mu X)^{2}}{4{(1-2X)}^2} =-\R_{\gamma\beta}v^\gamma v^\beta
\label{10}
\een
and the corresponding scalar expansion ($\tilde{\Theta}$) is \cite{Poisson,Das}
\ben 
\tilde{\Theta}=D_\a v^\a\equiv\frac{1}{\sqrt{-\G}}~\partial_\a(\sqrt{-\G}~v^\a)=3\frac{\dot{a}}{a}-\frac{v^\mu\partial_\mu X}{1-2X}.\nonumber\\
\label{11}
\een
where $\t=\nabla_\a v^\a$ is the usual scalar expansion, symmetric shear $\sigma_{\a\b}=\frac{1}{2}\Big(\nb_\b v_\a+\nb_\a v_\b\Big)-\frac{1}{3}\t h_{\a\b}$,  
antisymmetric rotation $\o_{\a\b}=\frac{1}{2}\Big(\nb_\b v_\a-\nb_\a v_\b\Big)$, $2\sigma^{2}=\sigma_{\a\b}\sigma^{\a\b}$, $2\o^{2}=\o_{\a\b}\o^{\a\b}$ and 
 $\nb_\b v_\a=\sigma_{\a\b}+\o_{\a\b}+\frac{1}{3}\t h_{\a\b}$
 with the three dimensional hypersurface metric $h_{\a\b}=g_{\a\b}-v_\a v_\b$. In \cite{Das}, the authors employed a DBI type non-canonical Lagrangian represented as $\La(X,\phi)= 1-V\sqrt{1-2X}$, where $V(\phi)\equiv V$ is a constant potential that is significantly smaller than the kinetic energy of the K-essence scalar field. 
 Here, in Eq. (\ref{11}), the background gravitational metric is assumed to be flat FLRW.
 
 Note that Eq. (\ref{10}) is the Raychaudhuri equation in the new geometry when $\phi$ is present. This is evident from equations (\ref{10}) and (\ref{11}) that this modified RE is form invariant when $v^{\mu}\partial_{\mu}X=0$ \cite{Ray,Kar,Poisson}. The modified scalar expansion reduces to its original form. It is important to note that the Ricci tensor ($\R_{\gamma\beta}$) on the right side of the modified RE (\ref{10}) differs from the original Ricci tensor ($R_{\gamma\beta}$) which defines a {\it different} geometry. This is as per the dynamical behavior of spacetime. However, we know that the Raychaudhuri equation is an identity (independent of any gravitational theories, created geometrically) and therefore form invariance of RE should be a logical consequence in any geometry defined by the Ricci tensor. We now show that $v^{\mu}\partial_{\mu}X=0$ admits non-trivial solutions for $\phi$ that ensure the recovery of the form of the original Raychaudhuri equation. Moreover, these solutions of $\phi$ are similar to inflatons, which are cosmologically significant in the early universe. Note that we have obtained cosmologically relevant solutions without taking direct recourse to the Friedmann equations, which is the usual practice.
 
 \section{ Solutions:} Considering the K-essence scalar field as inhomogeneous and isotropic, we define $\phi(x^{i})\equiv \phi(r,t)$ with a flat FLRW type background gravitational metric, $X$ is
\ben
X=-\frac{1}{2}\phi_{t}^2+\frac{1}{2a^2}\phi_{r}^2
\label{12}
\een
where $a=a(t)$ represents the scale factor and the metric signature is $(-,+,+,+)$. Taking $v^\mu$ as $(1,1,0,0)$, we have 
\ben
v^\mu\partial_\mu X=-\phi_{t}\phi_{tt}-\frac{\dot{a}}{a}\frac{1}{a^2}\phi_{r}^2+\frac{\phi_{r}\phi_{rt}}{a^2}-\phi_{t}\phi_{tr}+\frac{1}{a^2}\phi_{r}\phi_{rr}.\nonumber\\
\label{13}
\een
where $\phi_{t}=\frac{\partial\phi}{\partial t}$, $\phi_{r}=\frac{\partial\phi}{\partial r}$, $\phi_{tr}=\frac{\partial^{2}\phi}{\partial t \partial r}=\phi_{rt}$, $\phi_{tt}=\frac{\partial^{2}\phi}{\partial t^{2}}$, $\phi_{rr}=\frac{\partial^{2}\phi}{\partial r^{2}}$ and $\dot{a}=\frac{\partial a}{\partial t}$. The four vectors, $v^{\mu}$ may be expressed as $(1,1,0,0)$ for an inhomogeneous field. However, this selection is for a particular location in spacetime and may not be consistent elsewhere unless this specific choice is justified. $v^{\mu}=(1,1,0,0)$ might be seen as a specific instance in this situation, however, it does not serve as the overall answer for an inhomogeneous field. Furthermore, our study focuses solely on the invariance property of the Raychaudhuri equation when scalar fields are present without addressing any specific cosmological situation. Therefore, obtaining any result that may have any cosmological significance is a bonus. To solve $v^{\mu}\partial_{\mu}X=0$ take $\phi$ as 
\ben
\phi(r,t)=\phi_{1}(r)+\phi_{2}(t).
\label{14}
\een
Then Eq. (\ref{13}) becomes
\ben
a^2\dot{\phi}_2 \ddot{\phi}_2+\Big(\frac{\dot{a}}{a}\Big)(\phi'_{1})^2-\phi'_{1} \phi''_{1}=0
\label{15}
\een
where $'dot'$ means time derivative and $'prime'$ means space derivative. In equation (\ref{15}) there is a space and time derivative in the second term. To simplify this we take the scale factor as defined by \cite{Peebles2,Liddle,Weinberg, Mukhanov}:  
\ben a(t)=e^{H_0 t}.
\label{16} 
\een 
$H_{0}$ represents the positive constant value of Hubble parameter which may vary in different epochs. The specific scale factor can be utilized to describe the acceleration of the universe in later periods when it is primarily influenced by dark energy \cite{Weinberg} as well as during inflation in the early universe \cite{Peebles2, Liddle, Mukhanov}.
Basically during inflation the time derivative of the scale factor ($da(t)/dt$) was positive, in this case which emerges as $\dot{a}/a=H_{0}$. Therefore, the equation (\ref{15}) becomes
\ben
[H_{0}(\phi'_{1})^2-\phi'_{1} \phi''_{1}]=-e^{2H_{0}t}\dot{\phi}_2 \ddot{\phi}_2=w^2
\label{17}
\een
where $w^2~(\geq 0)$ is the separation constant.\\

{\it Case-I: $w^2=0$:} From Eq. (\ref{17}) we get the solutions as ($e^{2H_{0}t}\neq 0$):

\ben
\text{(i)}~~\phi(r,t)= constant = A+D;
\label{18}\\
\text{or, (ii)}~~\phi(r,t)\equiv \phi(t) =Bt+C+D;
\label{19}\\
\text{or, (iii)}~~\phi(r,t)=A+\frac{1}{H_{0}}e^{H_{0}r+E};
\label{20}\\
\text{or, (iv)}~~\phi(r,t)=Bt+C+\frac{1}{H_{0}}e^{H_{0}r+E};
\label{21}
\een
where $A,~B,~C,~D~\text{and}~E$ are the arbitrary integration constants. In the first example, $\phi(r,t)$ has four distinct kinds of solutions, and $v^{\mu}\partial_{\mu}X$ is zero for all of them. The modified RE (\ref{10}) reduces to the standard RE for these four categories of scalar fields. The constant scalar field is the first of the four solutions (\ref{18}), the homogeneous field is the second (\ref{19}), and the inhomogeneous scalar fields are the last two (\ref{20} and \ref{21}).  Thus, it can be concluded that the modified RE (\ref{10}) and the scalar expansion (\ref{11}) remain invariant for both the homogeneous and inhomogeneous scalar field solutions given above. Note that the fourth solution (\ref{21}) is the general solution of Eq.(\ref{17}) for $w^{2}=0$, which incorporates both the space part ($r$) and the time part ($t$). For $t=0$ one can achieve the third solution (\ref{20}), for $r=0$, one would get the second solution (\ref{19}) and for both $t=0~\text{and}~r=0$ one would get the first solution (\ref{18}). Typically, scalar field models are employed to elucidate the inflationary phases of the universe, which refer to the rapid expansion of space within a short period. In our fourth solution (\ref{21}), it is evident that the scalar field exhibits exponential dependence on space and linear dependence on time, suggesting inflationary behavior \cite{Mukhanov}. \bigskip

{\it Case-II: $w^2\neq 0$:} 
From the Eq. (\ref{17})  the general solutions of the K-essence scalar field are: 
\ben
\phi^{\pm}(r,t)&&=\mp\frac{1}{2 H_0^{3/2}}\Bigg[2 \sqrt{w^2+e^{2 H_0 (r+G_1)}}\nonumber\\
&&-w \ln \Big(w+\sqrt{w^2+e^{2 H_0 (r+G_1)}}\Big)\nonumber\\
&&+w \ln \Big(H_0 \Big(-w+\sqrt{w^2+e^{2 H_0 (r+G_1)}}\Big)\Big)\Bigg]+G_2\nonumber\\
&&\mp\frac{1}{2H_{0}}\Bigg[2 \Big(\sqrt{\frac{w^2}{H_{0}} e^{-2H_{0} t}+F_{1}}\nonumber\\
&&-\sqrt{F_{1}} \tanh ^{-1}\Big(\frac{\sqrt{\frac{w^2}{H_{0}} e^{-2H_{0} t}+F_{1}}}{\sqrt{F_{1}}}\Big)\Big)\Bigg]+F_{2}.\nonumber\\
\label{22}
\een
where $G_{1},~G_{2},~F_{1}~\text{and}~F_{2}$ are the arbitrary integration constants. It's important to note that $\phi^{+}(r,t)$ corresponds to the positive root solution, while $\phi^{-}(r,t)$ corresponds to the negative root solution of Eq. (\ref{17}) considering $w^2\neq 0$.

Furthermore, we may evaluate specific solutions of Eq. (\ref{17}) by setting $F_{1}=0$ and $\phi'_{1}(r)=\frac{w^2}{H_{0}}$, then the solutions can be written as 
\ben
\phi^{\pm}(r,t)=\pm \sqrt{\frac{w^2}{H_{0}}}~r+ G_3\mp \frac{1}{H_{0}}\sqrt{\frac{w^2}{H_{0}}}e^{-H_{0}t}+F_{2}
\label{23}
\een
where $G_3$ is an integration constant. Both solutions (\ref{22}) and (\ref{23}) are inhomogeneous. Equation (\ref{10}) and the scalar expansion (\ref{11}) reduce to the standard Raychaudhuri equation for the aforementioned inhomogeneous solutions of $v^{\mu}\partial_{\mu}X=0$. It should be mentioned that the constant solution of the field Eq. (\ref{18}) and particular solution Eq. (\ref{23}) are not viable in our instance, since these solutions give $X$ (\ref{12}) to be zero, which is not acceptable in our interacting model.

\section{Inflaton fields:} 
We now show that our solutions (\ref{19}),  
(\ref{21}) and (\ref{22}) are similar to inflaton fields as described in Mukhanov \cite{Mukhanov}. 

{\it Case-I: Homogeneous field:} The inflationary phase of the cosmos can be characterized by a scalar field known as ``inflaton". In the ultra-hard equation of state the kinetic component of the scalar field $(\dot{\varphi}^{2})  >> V(\varphi)$, where $V(\varphi)$ is the potential term, and the equation of state is $p \approx +\rho$, where $ \rho$ represents the standard energy density of the universe. In the ultra-hard stage \cite{Mukhanov}
\ben
\varphi=const-\frac{1}{\sqrt{12\pi}}\ln{t}.
\label{24}
\een
This corresponds to $H^{2}\simeq \frac{1}{9t^{2}}$, which directly follows from $a\propto t^{1/3}$ and $\rho\propto a^{-6}$. This solution is exact for a massless scalar field. The field's trajectory rapidly ascends and eventually reaches the attractor. The range of initial conditions increase and result in an inflationary phase. Once the trajectory intersects the attractor at a point when it is flat, denoted as $|\varphi|>>1$, the subsequent solution elucidates a phase of accelerated expansion. Then the corresponding scalar field is 
\ben
\varphi(t)\simeq \varphi_i-\frac{m}{\sqrt{12\pi}}(t-t_i)\simeq \frac{m}{\sqrt{12\pi}} (t_f-t)
\label{25}
\een
where $t_{i}$ is the time when the expansion starts and $t_f$ is the time when the scalar field dies out. Our solution (\ref{19}) in our particular setting exhibits a comparable linearly time-dependent solution for $\phi$. Hence, the K-essence scalar field that maintains the Raychaudhuri equation's invariance is similar to an ``inflaton" field.

{\it Case-II: Inhomogeneous field:} 

Next, we  show that our inhomogeneous solution (\ref{22}) is similar 
 to the inflaton solutions Eq.s (8.65) and (8.66) of Mukhanov \cite{Mukhanov} under specific circumstances using DBI type Lagrangian.

Considering a  DBI type Lagrangian as $p(X)=\La(X)=1-V\sqrt{1-2X}$ and  $B(X)=4\pi C \dot{\varphi}_{0}\sqrt{\frac{p_{X}}{c_s}}\cong D \Big(1+X+\frac{3}{2}X^2\Big)$
where $D=4\pi C \sqrt{V}\dot{\varphi}_0$. Again, considering $A(t)=k\int \frac{c_s}{a} dt$, it can be expressed as $\exp{(iA(t))}=1+iA(t)-\frac{A^2(t)}{2!}+\frac{(iA(t))^3}{3!}-...$, for real and positive solutions, we can recast the Mukhanov's inhomogeneous inflaton fields (8.65) and (8.66) \cite{Mukhanov} as
\ben
\Phi\simeq D \Big(1+X+\frac{3}{2}X^2+\frac{5}{2}X^3\Big) \Big(1-\frac{A^2(t)}{2!}-\frac{A^4(t)}{4!}\Big)\nonumber\\
\label{30}
\een
\ben
\delta \varphi &&\simeq C\sqrt{V}\Big[-\frac{k}{a}\Big(A(t)-XA(t)-\frac{X^2}{2}A(t)+\frac{A^3(t)}{3!}\nonumber\\&&-X\frac{A^3(t)}{3!}-\frac{X^2A^3(t)}{3!}\Big)+H\Big(1-\frac{A^2(t)}{2!}-\frac{A^4(t)}{4!}\Big)\Big]\nonumber\\
\label{31}
\een

Our general solution, (compare with (\ref{22})), should now be taken into consideration. The integration constants can be considered as zero for simplicity i.e., $G_1=G_2=F_1=F_2=0$. We also assume $w<<e^{2H_{0}r}$. Then, from (\ref{22}) the positive solution becomes,
\ben
\phi(r,t)=-\frac{1}{2H_{0}^{3/2}}\Big[2e^{H_{0}r}+w\ln{H_{0}}+2we^{-H_{0}t}\Big].
\label{32}
\een

Using (\ref{32}) and (\ref{12}), we express the radial part ($e^{H_{0}r}$) of (\ref{32}) in terms of $X$, we can recast the Eq. (\ref{32}) as
\ben
\phi=\mathcal{D}\Big[1+\mathcal{A}(t)X-\mathcal{A}^2(t) X^2+...+\frac{\ln{H_{0}}}{2}+\frac{1}{a^2}\Big]
\label{33}
\een
where $\mathcal{D}=-\frac{w}{H_{0}^{3/2}}$ and $\mathcal{A}(t)=\frac{H_{0}a^2}{w^2}$. So far, we have expressed our general solution of field (\ref{22}) in terms of $X$ and  $\mathcal{A}(t)$, while ensuring that $v^{\mu}\partial_{\mu}X=0$, maintaining the form invariance of the  Raychaudhuri equation (\ref{10}) without considering the curvature effect, namely the Ricci tensor of gravity. The solution of the field $\phi$ (\ref{33}) carries some resemblance to the inhomogeneous inflation fields (\ref{30}) and (\ref{31}) under certain conditions for the DBI type action mentioned earlier. 


Thus, the Raychaudhuri equation becomes relevant when considering primordial inhomogeneities within the framework of inflationary cosmology. In the inflationary scenario, the universe experiences a rapid exponential expansion, effectively smoothing out any inhomogeneities and anisotropies. 

However, during inflation, small inhomogeneities \cite{Mukhanov} can still remain that eventually develop into the large-scale structures we observe in the present-day universe \cite{Sarkar,Krasinski,Peebles,Secrest}. 
So that the Raychaudhuri equation may be employed to elucidate the origins of the primordial inhomogeneities via fields associated with inflation.

Alternatively, our particular inhomogeneous solutions (\ref{21}), (\ref{22}), and (\ref{32}) of the K-essence scalar fields derived from the invariant Raychaudhuri equation (\ref{10}) can be relevant when studying inhomogeneous cosmology which can be mapped on to the Lemaitre-Tolman-Bondi (LTB) metric \cite{Lemaitre, Tolman, Bondi, Bolejko, Enqvist}. In a different context, Gangopadhyay et al. \cite{dg3} developed the scaling relation for inhomogeneous K-essence scalar fields 
 $\phi(r,t)=\phi_{1}(r) + \phi_{2}(t)$ 
 (\ref{14}) in the context of the LTB metric.


\section{ Concluding remarks:} 
We have shown that the modified  Raychaudhuri equation (\ref{10}) reverts to its original form in the presence of both homogeneous and inhomogeneous scalar fields which are similar to inflatons. This is conceivable because the Raychaudhuri equation is based on geometry and not on any specific theory of gravitation. Our solutions based on the Raychaudhuri equation admit an inhomogeneous universe as well as a homogeneous universe. 
So the  Raychaudhuri equation is compatible with both homogeneous and inhomogeneous universe. We recall that there exists another interesting form invariance in cosmology, {\it viz.}, form invariance of the metric tensor, which leads to the existence of killing vectors.

Sarkar \cite{Sarkar} has provided a comprehensive explanation of inhomogeneous cosmology \cite{Krasinski}, supported by empirical data. In this article, the author elucidated the need of a novel kind of structure development for inhomogeneous and anisotrpic universes, as discussed by Krasinski \cite{Krasinski}. However, these models have not been extensively investigated, particularly in terms of their implications for structure formation, in comparison to the FLRW models. Recent observations \cite{Peebles, Secrest, Keenan, Celerier} indicate that the cosmos is inhomogeneous. This observational evidence is in sync with our results obtained from the equation regarding inhomogeneous scalar fields. Investigations for homogeneous scalar fields in light of the Raychaudhuri equation have been done in  \cite{Das}.\\

{\bf Acknowledgement:}
A.P. and G.M. acknowledge the DSTB, Government of West Bengal, India for financial support through Grant Nos. 856(Sanc.)/STBT-11012(26)/6/2021-ST SEC dated 3rd November 2023.  

{\bf Conflicts of interest:} The authors declare no conflicts of interest.

{\bf Data availability:} There is no associated data with this article, and as such, no new data was generated or analyzed in support of this research.

\end{document}